\documentclass[reprint,superscriptaddress,pre
]{revtex4-2}
\usepackage{amsmath,amssymb,bm}
\usepackage{natbib}
\usepackage{graphicx}
\usepackage{dcolumn}
\usepackage{hyperref}
\usepackage{changes}

\begin{document}

\title{Shear-Induced Wobbling and Motility Suppression in Swimming Bacteria}

\author{Wei Feng}
\author{Fanglong Dang}
\author{Hao Luo}
\affiliation{School of Physics, Northwest University, 710127, Xi'An, China}
\author{Alan C. H. Tsang}
\affiliation{Department of Mechanical Engineering, The University of Hong Kong, Hong Kong, China}
\author{Yanan Liu}
\email{yanan.liu@nwu.edu.cn}
\author{Guangyin Jing}
\email{jing@nwu.edu.cn}
\affiliation{School of Physics, Northwest University, 710127, Xi'An, China}

\begin{abstract}
The intricate wobbling motion of flagellated bacteria, characterized by the periodic precession of the cell body, is a determinant factor in their motility and navigation within complex fluid environments. While well-studied in quiescent fluids, bacterial wobbling under ubiquitous flow conditions remains unexplored. In this work, we investigate the wobbling dynamics of \textit{Escherichia coli} swimming near surfaces under steady shear flow. Our experiments reveal that the wobbling amplitude intensifies with flow strength before reaching a plateau, with this amplification exhibiting a strong dependence on the swimming orientation relative to the flow direction. It turns out that the enhanced wobbling remains governed by the misalignment between the cell body and the flagellar bundle. Furthermore, we observe that the wobbling frequency increases monotonically with flow strength, and that shorter bacteria exhibit more pronounced variations in both amplitude and frequency. By linking the wobbling motion to the intrinsic body-flagella misalignment, we attribute the flow-enhanced precession to a combination of shear- and chirality-induced torques acting on the flexible flagellar hook. This mechanical coupling ultimately suppresses the net migration velocity as the flow rate increases. These findings elucidate the elastohydrodynamic mechanisms by which shear flow modifies bacterial locomotion near surfaces, with implications for microbial transport in physiological and ecological environments.

\end{abstract}

\maketitle

\section{\label{sec:Intro}Introduction}

\begin{figure*}[t]
    \centering
    \includegraphics[width=\textwidth]{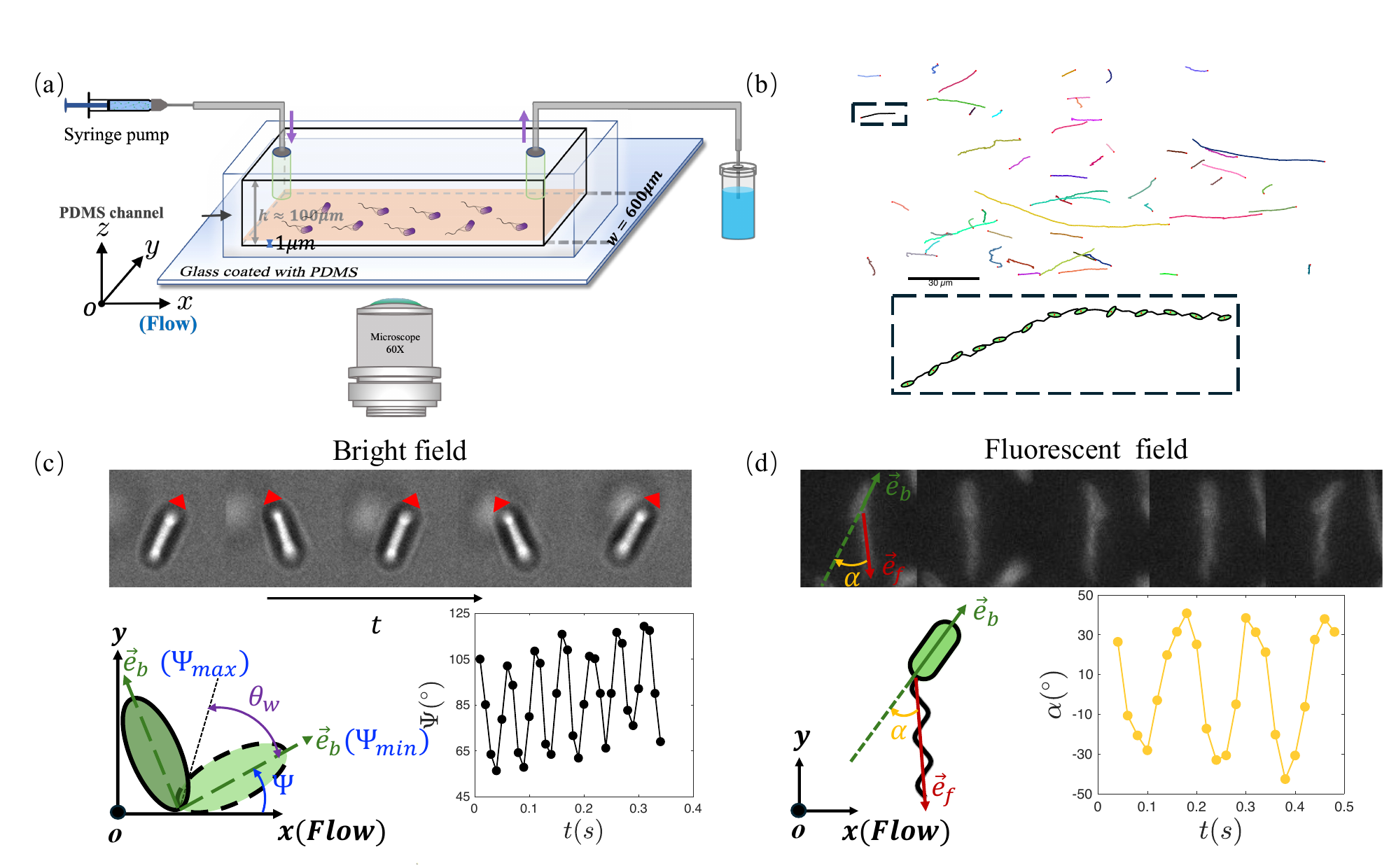} 
        \caption{\textbf{Experimental setup and kinematic characterization of bacterial wobbling.} 
        (a)Schematic of the microfluidic platform (height $H = 100~\mathrm{\mu m}$, width $W = 600~\mathrm{\mu m}$). A syringe pump maintains a controlled shear flow of the bacterial suspension. 
        (b) Typical zigzag trajectory of an \textit{E. coli} cell swimming in a quiescent motility buffer, captured $1~\mathrm{\mu m}$ above the bottom surface. Inset: High-magnification view of the trajectory segments. The mean cell body length is approximately $2~\mathrm{\mu m}$.
        (c) Definition of the wobbling angle $\theta_{w}$, determined as half of the peak-to-peak amplitude of the time-evolving cell-body orientation $\Psi$, called as polar angle in $x-y$ plane. The lower-right panel displays a representative periodic time trace of $\Psi$, and $\theta_w = (\Psi_{\text{max}} - \Psi_{\text{min}})/2$.
        (d) Measurement of the off-axis angle $\alpha$, quantifying the misalignment between the cell body's major axis and the rotation axis of the flagellar bundle. The lower-right panel illustrates the temporal variation of $\alpha$ measured via dual fluorescent labeling.}
        \label{fig:setup}
\end{figure*}
\begin{figure*}[t]
    \includegraphics[width=0.9\textwidth]{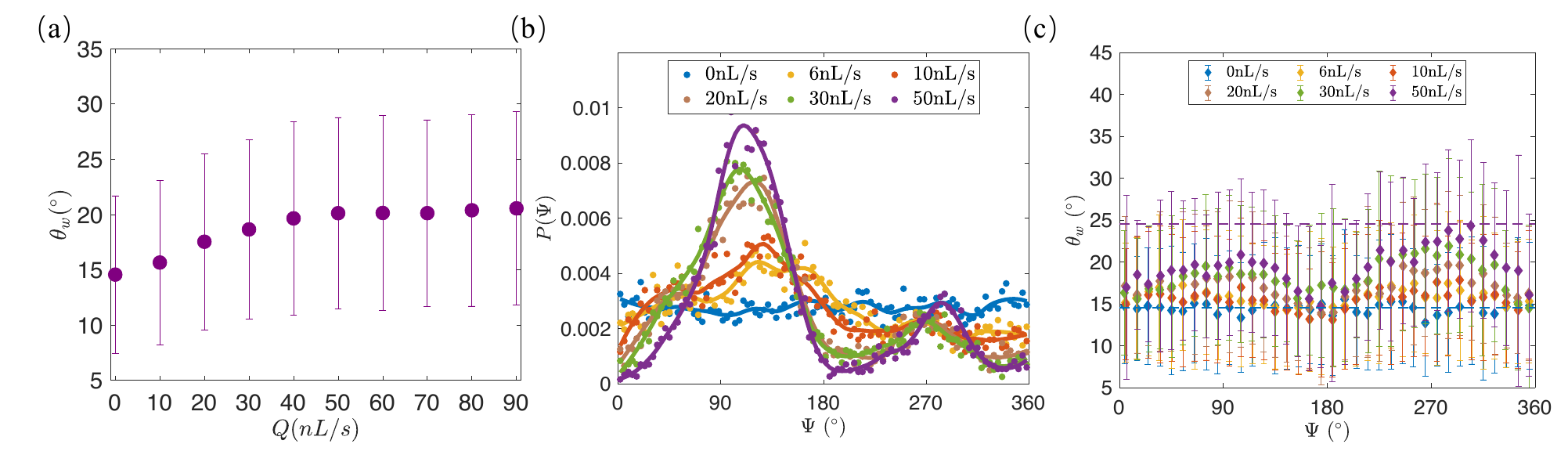} 
    \caption{\textbf{Dependence of bacterial wobbling on flow rate and cell body orientation.} 
        (a) Wobbling angle $\theta_w$ as a function of the flow rate $Q$. The amplitude of $\theta_w$ intensifies with increasing flow before reaching a plateau at high flow rates. Each data point represents an ensemble average of over 1000 cells; error bars denote the sample standard deviation.
        (b) Probability distribution of bacterial orientation $\Psi$ at various flow rates. Symbols represent bin centers, and solid lines are Gaussian-like fits provided as guides to the eye. The shift in orientation reflects a dynamic balance between wall-induced hydrodynamic torques, the weathervane effect, and chiral drifting torques, leading to a preference for upstream and transverse rheotactic alignment.
        (c) Orientation-dependent wobbling angle $\theta_w$ under different flow conditions (indicated by color). For a given flow rate, $\theta_w$ is minimized when the cell aligns with the streamlines and reaches its maximum when the orientation is perpendicular to the flow direction.
       }
    \label{fig:bacterial_wobbling_in_flow}
\end{figure*}
Optimize survival through diverse motility strategies, bacteria have evolved a wide array of morphologies, characterized by elongated cell bodies and specialized flagella that enable robust locomotion under fluctuating conditions \cite{lighthill1976flagellar, kysela2016diversity}. Such morphological evolution not only ensures locomotive efficiency but also dictates the emergence of distinct swimming gaits \cite{grognot2021more}.
As a simple form, monotrichous bacteria possessing a single polar flagellum adopt run-reverse or run-reverse-flick modes\cite{mitchell1996clustering,tian2022new,bubendorfer2014secondary}, whereas peritrichous bacteria such as \textit{Escherichia coli}, with flagella distributed over the entire cell surface, typically employ the run-tumble strategy alternating between directed swimming and random reorientation\cite{berg1972chemotaxis,ordal1975chemotaxis,cisneros2006reversal,tian2024swimming}.  

Among these gaits, the wobbling motion—a complex precession of the cell body—is a prominent feature of flagellated bacteria. 
This gait was first inferred from the zigzag trajectories of \textit{Spiroplasma}\cite{jennings1901significance}, and later confirmed as a widespread feature of microbial motility\cite{rikmenspoel1960cinematographic,keller1976swimming}. 
In 1972, three-dimensional tracking revealed wobbling in bacterial swimming\cite{berg1972chemotaxis}, and subsequent studies of magnetotactic bacteria further demonstrated their helical trajectories\cite{nogueiral1995study}. 
This behavior is fundamentally rooted in the off-axis alignment between the cell's major axis and the flagellar thrust, a broken symmetry that generates a torque perpendicular to the cell body and induces a precession-like motion significantly modulating their locomotion properties.
Swimming with precession rotation, bacteria follow helical paths in three-dimensional space that appear as zigzag trajectories in two-dimensional projections\cite{darnton2007torque,hyon2012wiggling}. 
Because wobbling is periodic, the instantaneous swimming direction oscillates, but the long-time swimming direction remains unchanged, which puts therefore the wobbling aside with  little impact on overall motility. 
However, later experiments showed that increasing polymer concentration reduces wobbling amplitude, straightens trajectories, and enhances swimming speed\cite{patteson2015running}, and then highlight the dominate the effect of viscosity and elasticity of the medium \cite{qu2020effects}. 
This debate has recently converged that experiments with colloidal suspensions demonstrated the same trend as in polymer solutions, namely reduced wobbling and increased speed with concentration, supporting the conclusion that wobbling is a key regulator of bacterial motility in complex fluids\cite{kamdar2022colloidal}.  

To date, the majority of research has focused on bacterial wobbling in quiescent fluids, where the gait is primarily determined by internal geometric asymmetries \cite{liu2014helical,constantino2016helical,mousavi2020wall,kamdar2022colloidal,hu2024flagellated}. 
In contrast, wobbling under flow remains poorly understood, despite the fact that flow is ubiquitous in bacterial habitats and profoundly alters swimming behavior\cite{lauga2016bacterial}.
In shear environments, flagellated bacteria experience non-equilibrium hydrodynamic forces and torques. 
More importantly, the chirality of the flagellar bundle induces a net transverse force that drives bacteria across streamlines\cite{marcos2012bacterial,marcos2009separation,jing2020chirality,zheng2023swimming}. 
Near surfaces, hydrodynamic and steric interactions further alter bacterial trajectories. As the shear rate increases, bacterial swimming evolves from circular motion to upstream migration, eventually leading to cross-stream reorientation\cite{nash2010run,hill2007hydrodynamic,kaya2012direct,figueroa2015living}. Hence, the questions arise how do flow-induced loads and chirality-governed forces coupled regulate the wobbling gait and body-flagella misalignment, and how does this modified wobbling, in turn, dictate the net migration speed of the bacteria.

In this work, we experimentally investigate the wobbling motion of flagellar bacteria in flat microfluidic channels under controlled shear flow, tracking the temporal evolution of the cell body orientation.
Bright-field imaging enables quantification of wobbling amplitude and frequency, while dual fluorescent labeling of the cell body and flagella allows measurements of the misalignment angle between the two axes. 
By analyzing the dependence of wobbling angle, frequency, and misalignment on flow strength, we characterize how shear flow regulates bacterial wobbling. 
We further investigate the modulating effects of swimming orientation and cell body length on the wobbling gait under flow. 
Subsequently, we examine the impact of wobbling on average swimming speed and compare the results with cases in quiescent simple and complex fluids. 
Finally, by using the resistive force theory framework, we link the wobbling motion to the intrinsic misalignment between body and rotating flagella.
Our analysis attributes the enhanced wobbling to a combination of shear- and chirality-induced torques acting on the flexible hook, which ultimately suppresses the swimming velocity as the flow rate increases.
Together, these results provide evidence and understanding to the coupling among external flow, bacterial morphology, and swimming gaits, and insights into microbial locomotion in realistic fluid environments.

\section{\label{sec:Result}Results}
\subsection*{Increased wobbling amplitude in flow}

To investigate the effect of shear flow on bacterial wobbling, we employed bright-field microscopy to observe and record the periodic oscillation of bacterial cell bodies at a height of approximately $1~\mathrm{\mu m}$ above the surface of a microfluidic channel (Fig. \ref{fig:setup}\textbf{a}). By binarizing the bacterial images, the cell body boundaries could be clearly distinguished, enabling us to extract the time evolution of the cell-body orientation $\Psi$ and the wobbling amplitude, defined as the wobbling angle $\theta_w = (\Psi_{\text{max}} - \Psi_{\text{min}})/2$, as shown in Fig. \ref{fig:setup} (\textbf{c}). As the flow rate increased, the wobbling angle gradually enlarged but eventually reached a plateau of approximately $20^\circ$ once the flow rate exceeded 40 nL/s, as shown in Fig. \ref{fig:bacterial_wobbling_in_flow}\textbf{a}. Each data point in the figure represents an average over more than 1000 cells, and the error bars denote the sample standard deviation. It is evident that, irrespective of the presence or absence of flow, the variation in wobbling angles among individual cells remains substantial.

Previous studies have shown that bacterial locomotion and orientation in flow result from the coupling of multiple effects, particularly near interfaces. These include periodic rotation induced by the coupling of shear flow with the elongated cell body\cite{rusconi2014bacterial, junot2019swimming}, cross-stream migration arising from the interaction between helical flagellar and shear, and in the vicinity of surfaces, circular and upstream swimming induced by hydrodynamic interactions with the boundary. Consequently, the cell-body orientation under different flow rates directly governs the forces acting on the bacterium and thereby its wobbling behavior. To examine this, we analyzed the distribution of bacterial orientations at various flow rates, as illustrated in Fig. \ref{fig:bacterial_wobbling_in_flow}(\textbf{b}). In a quiescent fluid, near the interface, the opposite rotations of the bacterial head and tail generate a hydrodynamic torque perpendicular to the surface, leading to an isotropic orientation distribution.

\begin{figure*}[t]
    \centering
    \includegraphics[width=0.9\textwidth]{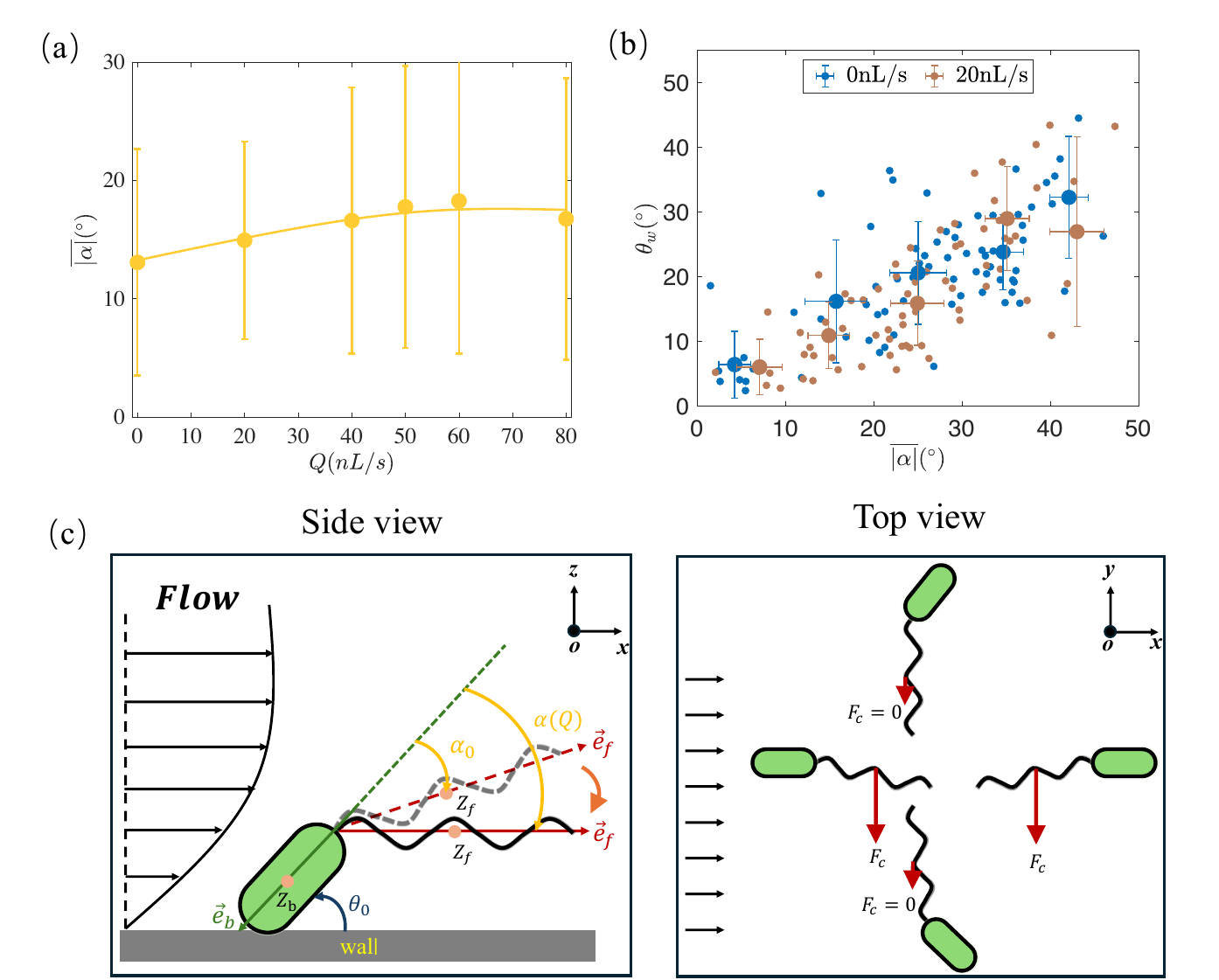}
        \caption{\textbf{Correlation between bacterial wobbling and body-flagella misalignment in flow.} 
        (a)Mean off-axis angle $\alpha$ as a function of the flow rate $Q$. The trend mirrors that of the wobbling angle: $\alpha$ initially increases with $Q$ before reaching a plateau at a critical shear rate of approximately $40\,\mathrm{s^{-1}}$. 
        (b)Similar trends between off-axis angle and wobbling angle are observed under both static and 20 nL/s flow conditions, with minimal differences at corresponding off-axis angles.Each small dot represents the off-axis angle and wobbling angle measured from an individual bacterium. The blue dots correspond to the off-axis and wobbling angles of bacteria in a quiescent environment, while the brown dots correspond to those at a flow rate of 20 nl/s. The larger dots represent the mean values within the corresponding intervals, and the error bars indicate the standard deviation of the bacterial wobbling angle within each interval.(c)Schematic of the force and torque driving hook deformation. In a simplified upstream-swimming configuration, the cell exhibits a "nose-down" tilt $\theta_0$ while the flagellar bundle is initially misaligned by $\alpha_0$. The coupling between the shear flow and the helical flagellum generates a shear torque and a chirality-induced transverse force $F_c$. These external loads are balanced by the elastic restoring torque of the hook, inducing a dynamic increase in the off-axis angle $\alpha$ as the flagellum reorients relative to the flow vorticity.}
        \label{fig:Alpha-Theta-flow}
\end{figure*}

As the flow rate increases, the weathervane torque arising from the coupling between the surface and the shear flow, thereby promotes upstream swimming. Specifically, as shown in Fig.\ref{fig:bacterial_wobbling_in_flow}(\textbf{b}), at 6 nL/s, bacterial orientations concentrate within the range of $\Psi=90^\circ \sim 180^\circ$. With further increases in flow rate, the drifting torque becomes dominant, driving the bacterial orientation to deviate further from the streamlines. Once the flow rate exceeds approximately 30 nL/s, the effects of the weathervane torque and the drifting torque reach a balance, and the orientation distribution exhibits a peak near $\Psi=135^\circ$. Given that wobbling is essentially manifested as periodic variations of cell-body orientation, this raises a key question: does flow-induced reorientation influence the amplitude of bacterial wobbling?

To address this, we quantified the wobbling angle $\theta_w$ of bacteria at different orientations under various flow rates. As shown in Fig. \ref{fig:bacterial_wobbling_in_flow}\textbf{c}, in the absence of flow, the wobbling angle remains nearly uniform across all orientations, around $\theta_w \sim 15^\circ$ indicating that torques arising from surface effect contribute negligibly to wobbling. Under flow, however, the wobbling amplitude exhibits pronounced orientation-dependent variations at the same flow rate, and these variations become more pronounced with increasing flow. In particular, cells aligned parallel to the flow direction display the smallest wobbling amplitude, whereas those oriented perpendicular to the flow exhibit the largest amplitude. These imply that torques arising from drifting effects and the weathervane mechanism play a crucial regulatory role in modulating bacterial wobbling.

\subsection*{Body–flagella misalignment dominates flow-enhanced wobbling}

To test shear-induced enhancement of bacterial wobbling, we measure the off-axis angle of flagella. In quiescent fluid, bacterial wobbling originates from the misalignment between the axes of cell body and the flagellar bundle. The wobbling angle $\theta_{w}$ depends on the off-axis angle $\alpha$, i.e., $\theta_{w} \propto \alpha$. Within the approximate range $0^\circ \lesssim \alpha \lesssim 20^\circ$, $\theta_{w}$ increases approximately linearly with $\alpha$. We therefore further examined the effect of shear flow on the off-axis angle. By fluorescently labeling both the flagellar bundle and the cell body, we measured the angle between them at different flow rates. As shown in Fig.\ref{fig:Alpha-Theta-flow}(\textbf{a}), the average off-axis angle $\alpha$ initially increases approximately linearly with flow rate and reaches a plateau at 40 nl/s. This trend is consistent with the variation of $\theta_w$  under flow. 
We therefore analyzed the relationship between the off-axis angle, $\alpha$, and the wobbling angle, 
$\theta_w$, under both quiescent conditions and a flow rate of 20 nl/s. As shown in Fig.\ref{fig:Alpha-Theta-flow}(\textbf{b}), the linear correlation between $\alpha$  and $\theta_w$ is nearly identical under both conditions within the measured range 
$0^\circ \lesssim \alpha \lesssim 50^\circ$. 
This observation indicates that the variations in wobbling amplitude, $\theta_w$, across different cell-body orientations are entirely attributable to changes in the off-axis angle $\alpha$.

\begin{figure*}[t]
    \includegraphics[width=0.9\textwidth]{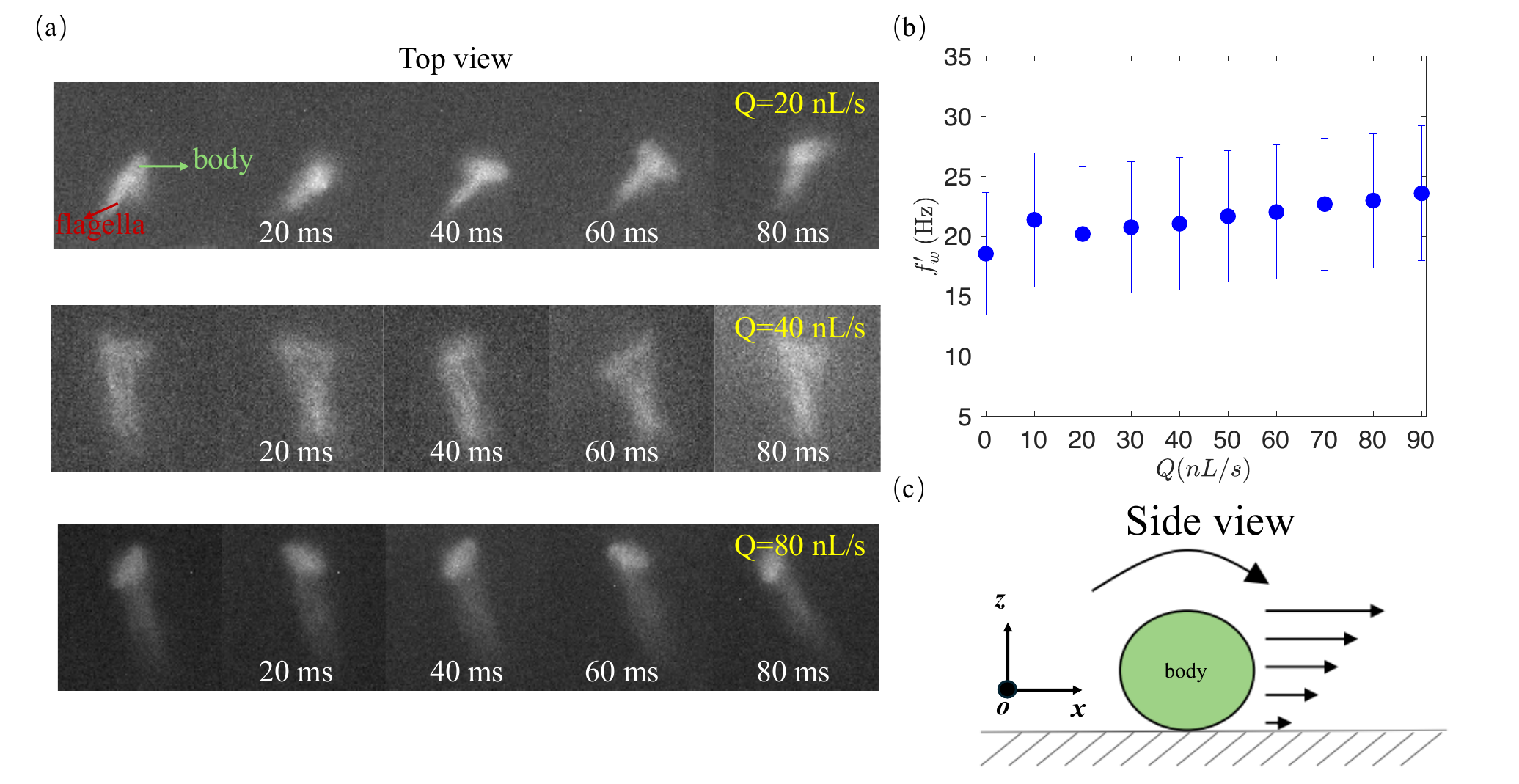}
        \caption{\textbf{Modulation of wobbling frequency by wall-induced rolling in shear flow.} 
        (a)Representative time-lapse snapshots (top view) of a bacterium undergoing rolling motion along the surface at various flow rates.
        (b)Apparent wobbling frequency $f^{'}_{w}$as a function of the flow rate $Q$, showing a monotonic increase with flow strength, in contrast to the decrease of pure wobbling with $\alpha$ (flow-induced) without rolling. 
        (c)Schematic illustration of the frequency compensation mechanism. While the increasing off-axis angle $\alpha$ induced by flow would theoretically lead to a decay in the intrinsic wobbling frequency, the no-slip boundary condition at the wall generates a shear-induced rolling motion. This rolling component superimposes onto the precessional dynamics, effectively compensating for the frequency decay and resulting in the observed rise in the total wobbling frequency with increasing flow rate.}
        \label{fig:WobblingWithRolling}
\end{figure*}
Early numerical simulations based on boundary element methods solved the Stokes equations afor single-flagellated cells near a surface suggested a tendency toward negative pitch angles (i.e., head-up)\cite{giacche2010hydrodynamic,pimponi2016hydrodynamics}. In contrast, subsequent experimental observations revealed that bacteria swimming near interfaces predominantly exhibited positive pitch angles (i.e., head-down)\cite{bianchi20193d,bianchi2017holographic}, while in some cases both positive and negative pitch angles were observed to coexist\cite{liu2024bacterial}. It has also been demonstrated that the pitch angle decreases with increasing flow rate and eventually reaches a plateau\cite{mathijssen2019oscillatory}. More recently, theoretical studies incorporating short-range thermodynamic interactions between bacteria and surfaces have effectively accounted for the deviations between simulation and experimental \cite{leishangthem2024thermodynamic}.It is still challenging to precisely measure how flow influences bacterial orientation and pitch angle during near-surface swimming, thereby modulating the off-axis angle and wobbling angle.

\begin{figure*}[t]
    \includegraphics[width=0.9\textwidth]{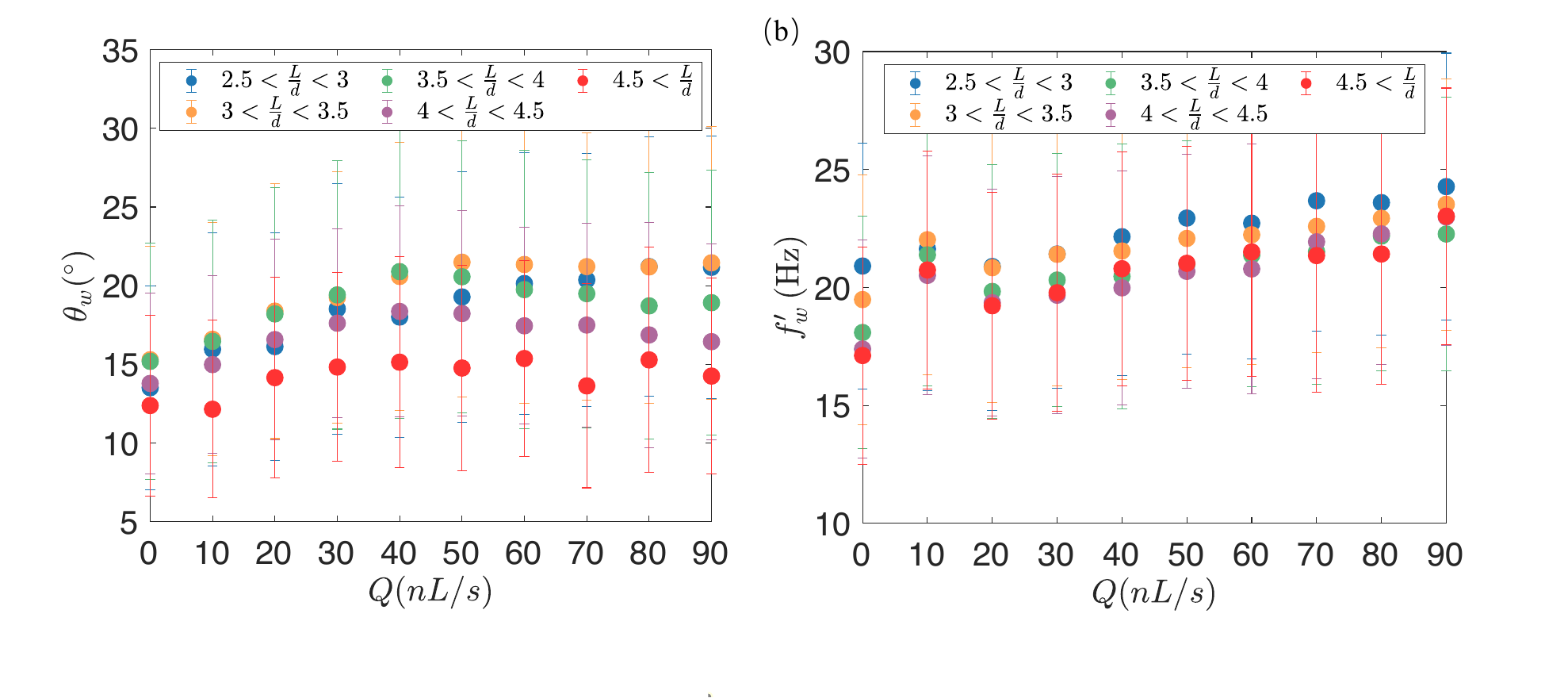} 
        \caption{\textbf{Variation of bacterial wobbling with flow for different body length.} 
        (a)Different colors represent different body length. With an increase in bacterial cell body length, the variation in wobbling angle in response to changes in the flow decreases. (b) With an increase in bacterial cell body length, the variation in wobbling frequency in response to changes in the flow  also decreases.}
        \label{fig:wobbling for different body length} 
\end{figure*}

As the shear rate increases, the bacterial cell body gradually aligns more parallel to the surface, leading to a decrease in the pitch angle, which eventually stabilizes. Concomitant with changes in pitch angle, the relative distance between the flagellar bundle and the surface also varies, thereby modifying the frictional force exerted by the boundary on the bundle. This, in turn, alters the torque applied by the flow on the flagellar bundle, inducing deformation of the elastic flagellar hook and consequently enhancing the wobbling.

In the shear flow, the flagellar bundle also experiences a transverse drifting along the vorticity direction\cite{marcos2012bacterial,jing2020chirality}. In addition, due to swimming in close proximity to the interface, both the cell body and the flagellar bundle are subjected to torques induced by the weathervane effect (see Fig. \ref{fig:Alpha-Theta-flow}(\textbf{c})). Because of morphological variations among individual bacteria, such as differences in cell-body and flagellar length, the resulting stable polar angle $\Psi$ differs across cells. Consequently, distinct wobbling happens under the same flow rate, i.e., the more perpendicular the cell-body orientation is to the flow direction, and hence the larger the wobbling amplitude.
As the flow rate increases, the mean polar angle $\bar{\Psi}$ decreases from $180^\circ$ (perfectly upstream) to approximately $135^\circ$. Importantly, the critical shear rate at which the polar angle\cite{mathijssen2019oscillatory}, the flagellar-bundle centroid–wall distance, and the mean polar angle $\bar{\Psi}$ all converge to their respective steady-state values is consistent. Therefore, with increasing flow rate, the wobbling amplitude first increases and then reaches a plateau. It can be inferred that the effect of flow on bacterial wobbling is not exerted directly on the motion itself, but rather acts indirectly by first modulating the pitch angle, thereby altering the cell’s off-axis angle and consequently regulating the wobbling amplitude.

In addition to modulating the wobbling frequency by shear flow as the case in bulk flow, the wall surface also influences the apparent wobbling frequency\cite{bianchi2015polar} of bacteria due to rolling behavior with the anchoring on the wall. Fig. \ref{fig:WobblingWithRolling}(\textbf{a}) illustrates the morphology and orientation of the bacterial cell body and flagellar bundle during wobbling at different flow rates. As shown, the wobbling frequency increases monotonically with the strength of the shear flow. With increasing flow rate and shear, the bacterial trajectory evolves from near-circular motion in quiescent fluid to cross-stream migration. Consequently, the cell body orientation gradually aligns with the direction of flow vorticity. Under this condition, the bacterial wobbling dynamics are superimposed with a rolling motion induced by the shear vorticity, as schematically depicted in Fig. \ref{fig:WobblingWithRolling}(\textbf{c}). Since the vorticity increases with flow rate, the rolling frequency of the bacterial body also rises approximately linearly with the flow rate.

\subsection*{Bacterial shape changes wobbling motion}
The physical origin of wobbling lies in the partial transmission of driving torque into the precessional motion of the cell body via the flexible hook; consequently, the geometric dimensions of the cell body and flagellar bundle emerge as critical determinants of the wobbling dynamics. While the cell body diameter remains relatively constant across the population, its length exhibits significant variability. As shown in Fig. \ref{fig:wobbling for different body length}(\textbf{a}), at a given flow rate, shorter bacteria undergo more pronounced wobbling than their longer counterparts. Specifically, bacteria with a body length of approximately $5\,\mu\text{m}$ maintain a relatively stable wobble angle of $\theta_w \sim 12^\circ$, whereas shorter cells exhibit substantially larger angles that intensify with increasing flow rate. This intensification typically saturates at a shear rate of approximately $40\,\text{Hz}$ (corresponding to a flow rate of $40\,\text{nL/s}$), in Fig. \ref{fig:wobbling for different body length}(\textbf{b}). Interestingly, despite the differences in body dimensions, the wobbling frequency for all observed bacteria follows a universal linear dependence on the flow rate, spanning a range from $16\,\text{Hz}$ to $23\,\text{Hz}$.

The wobbling dynamics of bacteria are governed not only by external factors such as flow fields and colloidal interactions but also by intrinsic structural features\cite{clopes2021flagellar,kamdar2023multiflagellarity}, including cell-body length and flagellar length. It is reasonable to expect that the wobbling amplitude decreases with increasing cell-body aspect ratio, which is obvious by considering drag coefficients in the RFT theory on wobble and will be discussed in the later section. We confirm this consistence with the same trend under quiescent conditions (see Fig. \ref{fig:wobbling for different body length}(\textbf{a})). However, in natural environments, bacterial wobbling is in fact determined by the coupling between external perturbations and internal structural characteristics. To quantitatively assess the role of body length under flow, we compared the wobbling amplitude and frequency of bacteria with different body lengths at a fixed flow rate. The results show that, under shear, cells with larger aspect ratios exhibit a weaker response, to say, at the same flow rate, they display smaller wobbling amplitudes and lower frequencies.

\begin{figure*}[t]
    \includegraphics[width=0.9\textwidth]{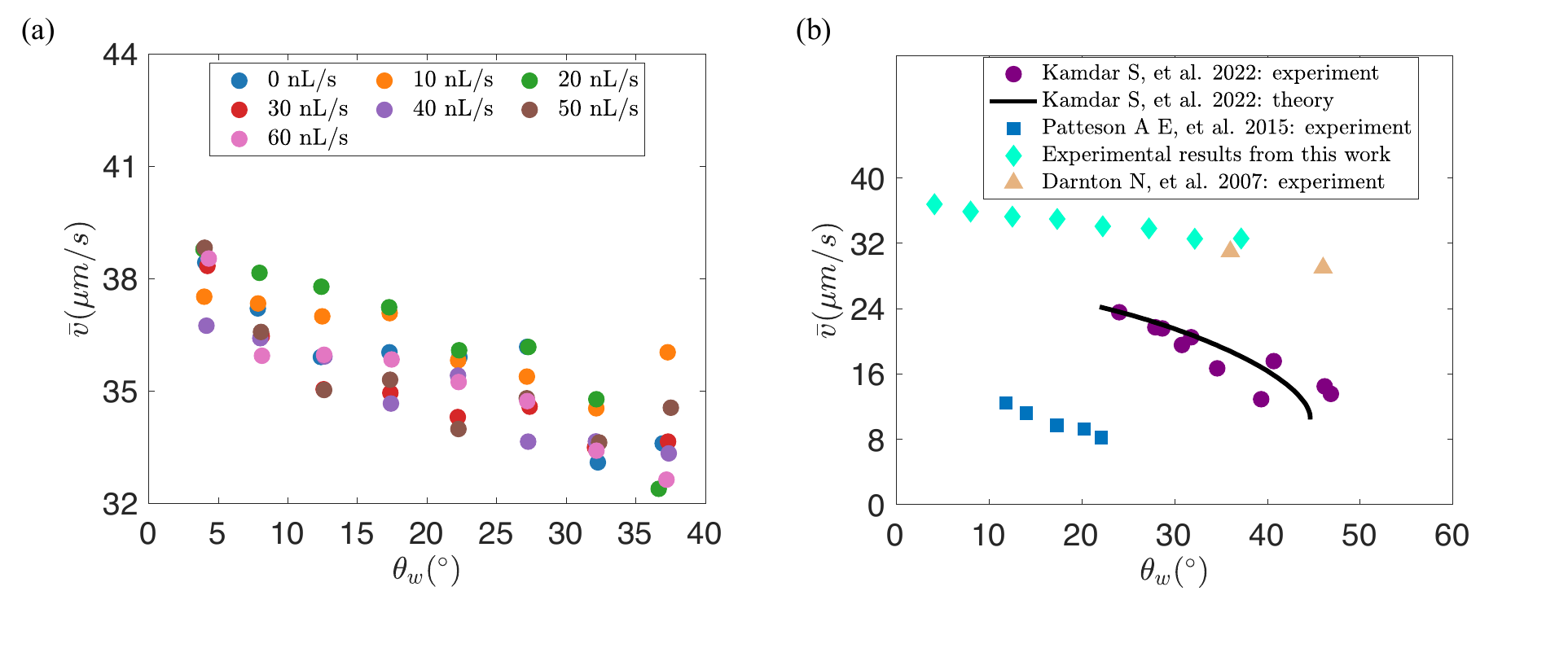}
        \caption{\textbf{Supression on swimming speed by enhanced wobbling in shear flow.} 
        (a)Wobbling reduces the average speed in a specific shear flow. As wobbling amplitude increasing, speed is decreasing.Different colors represent different flow rates. (b)The decreasing trend of speed with wobbling in the flow is compared with previous results. Different colors represent the results obtained by different researchers. The diamonds denote the averaged wobbling amplitude and swimming speed across different flow.}
        \label{fig:WobbleSpeed}
\end{figure*}

\subsection*{Wobbling suppression of swimming speed under shear flow}

Owing to the periodicity of wobbling, the long-time average swimming direction remains unaffected, which remind that wobbling has only a limited impact on overall bacterial motility. Later investigations \cite{kamdar2022colloidal}, however, revealed that such helical trajectories can substantially reduce the net displacement over a given time, lowering the mean swimming speed by as much as $80\%$. In contrast, under our experimental conditions—where both the observation region and bacterial strains differ from those used in previous studies—we observed a maximum reduction of only $15\%$ in swimming speed [see Fig. \ref{fig:WobbleSpeed}]. During wobbling, larger bending deformations of the flagellar hook correspond to larger wobbling angles. Nevertheless, a theoretical model assuming a constant motor torque output\cite{martinez2014flagellated} demonstrated that even an increase of the wobbling angle from $\theta_w\sim 20^\circ$ to $45^\circ$ alters the torque transmission by less than $7\%$. This result is consistent with our observations that hook bending–induced wobbling does not substantially affect the instantaneous swimming velocity.

Basically, as for the bacterial yawing typically it is assumed that the mean swimming speed remains constant with flow\cite{marcos2012bacterial}. Our experiments, however, demonstrate that external flow can substantially modulate the wobbling angle, implying that the mean swimming speed may also be influenced, even the self-propelled speed at the motor level. As bacteria are advected by streamlines in flow, directly measuring the intrinsic swimming velocity of individual cells is challenging. Following the common agreement, the mean swimming velocity in three dimensions is defined along the central axis of the helical trajectory, which coincides with the wobbling axis in two-dimensional projections. Since the drifting velocity in the transverse direction, $v_y$, arises solely from self-propulsion, we projected $v_y$ onto the wobbling axis to estimate the mean swimming speed. 

By this method, we measure that the mean swimming velocity is not constant under flow. At a fixed flow rate, the swimming speed decreases with increasing wobbling amplitude [see Fig. \ref{fig:WobbleSpeed}(\textbf{a})]. Furthermore, by averaging the data across different flow rates within the same wobbling amplitude intervals and comparing them with prior reports [Fig. \ref{fig:WobbleSpeed}(\textbf{b})], we conclude that the mean swimming speed decreases gradually as the wobbling amplitude increases. Although quantitative differences exist among various studies, our experimental data (denoted by diamond symbols) are consistent with the results  in Ref. [\cite{kamdar2022colloidal}]. Given that the same bacterial strain was used in both studies, the observed discrepancies are most likely attributable to strain-dependent differences in swimming speed.

\section{Discussion}

The experimental results clearly demonstrate the dependence of wobbling and motility on shear flow, revealing a robust correlation between the reduction in net migration velocity and the intensification of the wobbling gait. To elucidate the mechanisms underlying these observations, the subsequent discussion proceeds from the static configuration of the bacterial gait to dynamic feedback in flow, and from shear-induced torque to the chiral amplification of flagellar reorientation relative to the cell body, ultimately governing the wobbling precession. First, we introduce a general kinematic model based on Resistive Force Theory (RFT) to explore how the off-axis alignment between the cell body and flagellar axes serves as the physical origin of wobbling. Subsequently, the discussion focuses on the elastohydrodynamic response of the flagellar hook, analyzing its role as a central mediator that translates flow-induced shear torques into dynamic deformations of the off-axis angle. Finally, to address the transverse drift observed in our experiments, we incorporate asymmetric torques arising from flagellar chirality and rheotactic behavior. By examining this complex wobbling in shear flow, we aim to provide a mechanistic explanation for the intricate wobbling process and the coupling of bacterial motility with strong shear environments.

\subsection*{Resistive force theory for bacterial wobbling}

The bacteria has body with a length of $2l$, a radius of $r_b$, and a unit vector $\mathbf{e}_b$ defining the orientation of its major axis. \textit{Escherichia coli} has multiple flagella  randomly distributed on the cell surface, with each flagellum connected to an independent motor via a flexible hook, allowing them to rotate synchronously as a flagellar bundle to drive the overall motion of the bacterium. Here, the bundled flagella are simplified as a single helical flagellum with an axial length $L$, helix radius $R$, pitch $\lambda$. The cross-sectional radius of the slender flagellar filament is $a$, and the unit vector of the flagellar rotation axis is $\mathbf{e}_f$, pointing from the hook toward the distal end. In the laboratory frame, the angle between $\mathbf{e}_b$ and $\mathbf{e}_f$ breaks the overall symmetry of the bacterium, leading to
 a helical trajectory of the center of mass of a precessing bacterium in 3D space, on which the translational velocity vector $\mathbf{U}$ is directed tangentially at each position. Vector-averaging the instantaneous velocity over a complete precession period yields the axial average velocity $\mathbf{V}$, characterizing the bacterium motility.The angular velocity of the cell body is represented by $\bm{\Omega}$. The angular velocity of the helical flagellum relative to the cell body is $\bm{\omega}_m = \omega_m \mathbf{e}_f$, and the total angular velocity of the flagellum in the laboratory frame is $\bm{\Omega}_f = \bm{\Omega} + \bm{\omega}_m$. The swimming gait of the bacterium can be characterized by the \textit{off-axis angle} $\alpha$, the angle between the major axis of the cell body and the rotation axis of the flagellum, defined as $\cos \alpha = \mathbf{e}_b \cdot \mathbf{e}_f$, representing the geometric configuration of the bacterial body and flagellum, which may vary due to changes in configuration of the cell swimmer under different flow shears or external forces.
The major axis $\mathbf{e}_b$ and the precession axis $\mathbf{e}_p$ (the centerline of the helical trajectory) no longer coincide but form an angle known as the wobble angle, $\theta_w=\angle(\mathbf{e}_p, \mathbf{e}_b)$, describing the degree to which the body shakes or deviates from the central propulsion axis during its 3D precession.

At low Reynolds numbers, force and torque are related to velocities through the resistance matrices $\mathbf{A, B, C}$. In the absence of inertia, the total force and torque acting on the bacterium must be zero, leading to the following equation:
\begin{equation}\label{eq:TotalFT}
\begin{pmatrix} \mathbf{A}_b + \mathbf{A}_f & \mathbf{B}_f^T \\ \mathbf{B}_f & \mathbf{C}_b + \mathbf{C}_f \end{pmatrix} \begin{pmatrix} \bm{U} \\ \bm{\Omega} \end{pmatrix} =
\begin{pmatrix}
\mathbf{F}_{\text{prop}}\\
\mathbf{T}_{\text{prop}}
\end{pmatrix}
\end{equation}
The subscripts $b$ and $f$ denote the cell body and flagellar, respectively. The left side of the above equation represents the total force and torque experienced by the cell body and flagellum, while the right side represents the thrust and driving torque generated by the rotation of the flagellar motor. It indicates that the left side represents the resistance (force and torque) felt by the bacterium, while the right side arises from flagellar rotation and the continuous deformation/rotation of the flagellum relative to the body at an angular velocity $\bm{\omega}_f$. The active driving terms are $\bm{F}_{\text{prop}} = -\mathbf{B}_f^T \bm{\omega}_m$ and $\bm{T}_{\text{prop}} = -\mathbf{C}_f \bm{\omega}_m$. The cell body and flagellum form a specific configuration at a given off-axis angle $\alpha$, which determines a transformation matrix $\mathbf{Q}$. The flagellar contributions to matrices $\mathbf{A, B, C}$ must be transformed into the unified cell body coordinate system using this transformation matrix. Thus, all terms in the total force and torque equation (\ref{eq:TotalFT}) are established, and all quantities ultimately become functions of geometric parameters, physical parameters, and the rotation speed of the flagellum relative to the body $\omega_m$. The translational and rotational velocities of the bacterium can be obtained by solving Eq. (\ref{eq:TotalFT}). The solution for the angular velocity $\bm{\Omega}$ is:
\begin{equation}
\bm{\Omega} = (\mathbf{C} - \mathbf{B} \mathbf{A}^{-1} \mathbf{B}^T)^{-1} \left( \mathbf{T}_{\text{prop} } - \mathbf{B} \mathbf{A}^{-1} \mathbf{F}_{\text{prop}} \right)
\end{equation}
Then, with the angular velocity $\bm{\Omega}$, the expression for the translational velocity $\mathbf{U}$ is:
\begin{equation}
\mathbf{U} = \mathbf{A}^{-1} \left[ \mathbf{F}_{prop} - \mathbf{B}^T (\mathbf{C} - \mathbf{B} \mathbf{A}^{-1} \mathbf{B}^T)^{-1} (\mathbf{T}_{prop} - \mathbf{B} \mathbf{A}^{-1} \mathbf{F}_{prop}) \right]
\end{equation}

The geometric structural parameters of the bacterium determine the specific values of the resistance matrices $\mathbf{A, B, C}$, including the off-axis angle $\alpha$, which is the critical parameter for wobble motion. It alters the projection of the flagellar resistance elements in the cell body system through the rotation matrix $\mathbf{Q}(\alpha)$. When $\alpha$ increases, the thrust vector $\mathbf{F}_{prop}$ deviates from the main axis, causing energy to be wasted on lateral oscillations (wobble), resulting in a significant decrease in the axial swimming speed $U$. Regarding cell body size ($l, r_b$) and flagellar length ($L$), a larger body leads to a larger $\mathbf{A}_b$ (increased resistance) and slower speed. A longer flagellum increases the $\mathbf{B}$ term (thrust term) but also increases $\mathbf{A}_f$ (resistance term).
\begin{equation}
\mathbf{U}(\alpha) = - B_{zz}^f \omega_m 
\begin{pmatrix} 
\frac{\sin\alpha}{\mathcal{R}_x(\alpha)} \left[ 1 - \mathcal{C}(\alpha) + \mathcal{C}'(\alpha) \right] \\
0 \\
\frac{\cos\alpha}{\mathcal{R}_z(\alpha)} \left[ 1 - \mathcal{C}(\alpha) - \mathcal{C}'(\alpha) \right]
\end{pmatrix}
\label{eq:U_alpha}
\end{equation}
The instantaneous translational velocity $\mathbf{U}(\alpha)$ of the bacterium is intrinsically coupled to the off-axis angle $\alpha$ through the orientation-dependent resistance elements. As $\alpha$ increases due to the external shear torque, the redirection of the thrust vector $B_{zz}^f \omega_m$ leads to a non-linear modulation of both the axial migration and the transverse drift. Specifically, the anisotropy in drag coefficients $\mathcal{R}_{x,z}(\alpha)$ and the primary hydrodynamic correction $\mathcal{C}(\alpha)$ govern the deviation of the swimming path from the cell's major axis. The auxiliary term $\mathcal{C}'(\alpha)$ further accounts for the higher-order coupling between rotational and translational resistance that differs across the axial and transverse directions. Together, as shown in the \textit{suplementary materials Fig. S1 and S3}, these orientation-dependent factors dictate the net migration motility in complex shear flow.

In non-coaxial scenarios, the relationship between the bacterium's motion (instantaneous translational velocity $\mathbf{U}$ and angular velocity $\bm{\Omega}$) and the off-axis angle $\alpha$ is not a simple function. According to the definition of the wobble angle, $\bm{U}$ and $\bm{\Omega}$ are also associated with the wobble angle $\theta_w = \arccos (\mathbf{e}_b \cdot \mathbf{e}_p)$. Based on this calculation, we set different bacterial configurations by adjusting the off-axis angle $\alpha$ to examine changes in the wobble angle. The results show qualitatively that increasing the off-axis angle causes the wobble angle to increase monotonically and the migration speed $\mathbf{V}$ to decrease monotonically (Fig. S1). This is consistent with expectations and aligns well with our experimental results in Fig.(\ref{fig:bacterial_wobbling_in_flow}a) and Fig. (\ref{fig:wobbling for different body length}b) that the off-axis angle is approximately equal to the wobble angle, in Fig. (\ref{fig:Alpha-Theta-flow}b).

\subsection*{Hook anchoring and compliance modulating the driving torque}
The experimental results show that the measured swimming speeds $\sim 40 \mu$m/s are much larger than those calculated by theory (around $\sim 20 \mu$m/s) with setting of the similar motor rotation speed (Fig. S1), which reminds us that the flexibility and position of the flagellar hook are important factors. The \textit{E. coli} has multiple flagellar with individual hook as a flexible universal joint, thus the torque generated by body rotation cannot be fully transmitted to the mounted flagellum. We can simulate this effect by introducing an effective coupling coefficient $\gamma$ ($0 \le \gamma \le 1$): $\gamma = 1$ corresponds to complete rigidity, while $\gamma = 0$ means the rotation produces no thrust gain.
From the numerical calculation (Fig. S1) of the wobble angle by varying hook bending rigidity, a smaller $\gamma$ (hook softening) typically results in a smaller wobble angle because $\gamma$ effectively reduces the "lever arm efficiency" of the cell body's response to eccentric flagellar thrust. 

The flagellar hook plays a key role in both the magnitude and direction of bacterial swimming speed \cite{xie2011bacterial, son2013bacteria, sporing2018hook}. It has been verified that hardening the flagellar hook can prevent bundling, as a rigid hook tends to remain perpendicular to the cell membrane surface rather than bending \cite{brown2012flagellar, nguyen2018impacts, wu2025run}. Interestingly, it was reported that when the hook is soft, it can automatically bend to an appropriate stable angle during swimming regardless of its initial orientation. However, when the hook is very stiff, exceeding a certain critical stiffness, swimming stability is lost. At an intermediate stiffness, the flagellum can align well with the cell body axis, but surprisingly, the swimming speed does not change significantly \cite{shum2012effects}. Recent work has also confirmed that the flexibility of the hook in single-flagellated bacteria greatly changes their tumble process and residence time near walls, and the position of the hook is equally important \cite{tao2025flagellar}. Najafi et al. measured a monotonic decrease in swimming speed with increasing wobble angle: as the wobble angle rose from $5^\circ$ to $40^\circ$, the speed dropped from 35 $\mu$m/s to 15 $\mu$m/s \cite{najafi2019swimming}. On the other hand, bacteria exhibits various flagellar attachment positions, ranging from clustering near one end of the cell to being randomly distributed over the whole body \cite{brown2012flagellar, wu2025run}. As long as hook position $\mathbf{r}_h$ is away from the center of mass, the thrust generates a "flipping torque." In the wobble model by Hyon et al. \cite{hyon2012wiggling}, it is precisely the torque generated by this anchor offset that breaks rotational symmetry, leading to wobble.

\subsection*{Enhanced wobbling in shear flow}

In a long PDMS channel with square cross-section channel (height $H$, width $W$), to avoid side-wall effects, our observation window was located in the central region of the width ($y$-direction), on a height of 1 $\mu$m from the bottom wall. We can measure the cell body's center of mass undergoing wobble motion, appearing as a zigzag trajectory. Bacteria near the wall tend to swim upstream in shear flow and are not parallel to the wall, and instead the cell body tilts downward and the flagellum tilts upward at an angle $\theta_0 = 10^\circ$ \cite{mathijssen2019oscillatory}, a posture also known as "nose-down" \cite{leishangthem2024thermodynamic}. Since the bacterial head is close to the wall, while the tail (flagellum) is farther from the wall, making its $z$ larger, bacteria experiences a shear stress by flow. As the flow rate $Q$ increases, the hydrodynamic torque acting on the "high-position" flagellar bundle increases. On a flexible hook, this overcomes the hook's restoration force, leading to an increase in the flagellar off-axis angle $\alpha$, attempting to push the flagellum further outward. Also, the chiral flagellum experiences lift force perpendicular to the flow then reorients the bacteria with an additional torque. Therefore we are trying to incorporate the flow into the RFT model to understand the wobbling in flow in the following discussion. 

\subsubsection{Shear-torque-induced changes in the off-axis angle}
According to geometric relationships, if the cell body length is $2l$ and the nose-down tilt is $\theta_0$, then $\Delta z \approx (l + L/2) \sin(\theta_0)$, where $L$ is the length of the flagellar bundle. The flagellum experiences a flow velocity faster than the cell body: $\Delta u = \dot{\gamma} \cdot \Delta z \approx \dot{\gamma} \cdot (l + L/2) \sin(\theta_0)$.
We treat the flagellar bundle as a cantilever beam subjected to fluid drag, with a flexible hinge at its base. Shear flow generates an additional drag force on the flagellum. According to drag theory, for a point on the flagellum at a distance $s$ from the hook, the lateral viscous drag intensity (force per unit length) at that point is $f_{\perp}(s) = \zeta_{\perp} \cdot \Delta u(s) = \zeta_{\perp} \cdot \dot{\gamma} \cdot (s + l) \sin \theta_0$, where $\zeta_{\perp}$ is the transverse drag coefficient of the flagellum. For a slender body, $\zeta_{\perp} \approx 4\pi\eta/[\ln(L/a) + 0.5]$, where $a$ is the flagellar radius. Since this force acts at a certain distance from the hook, it generates an additional bending torque $T_{flow}$ around the hook. $T_{flow}$ is obtained by integrating the force over the entire flagellum:
\begin{equation}\label{eq:ShearTorque}
T_{flow} = \int_{0}^{L} f_{\perp}(s) \cdot s \, ds = \zeta_{\perp} \cdot \dot{\gamma} \cdot \sin \theta_0 \cdot \left( \frac{L^3}{3} + \frac{lL^2}{2} \right)
\end{equation}
At a flow rate $Q \sim 100$ nL/s, this shear torque can reach the order of $10^{-16}$ N$\cdot$m. The hook, as an elastic element, has a deflection angle $\alpha$ that depends on the total torque it receives. In the presence of a flow field, the total torque is the vector sum of the intrinsic deviation torque $T_{motor}$ generated by the motor and the flow-induced torque $T_{flow}$. Assuming the hook is within a small deformation range, the restoration torque $T_{restore} = K_h \cdot \alpha$. Thus, $K_h \cdot \alpha = T_{motor} + T_{flow}$. Defining $T_{motor}$ as the term generating the initial off-axis angle $\alpha_0$, i.e., $K_h \alpha_0 = T_{motor}$,then,
\begin{equation}
\label{eq:OffAxisAngleFlow}
\alpha(Q) = \alpha_0 +\mathcal{K} \cdot \dot{\gamma}(Q) \cdot \sin(\theta_0)
\end{equation}
where $\mathcal{K}= C \cdot \eta \cdot L^2/K_h$ is the structural compliance parameter, encompassing fluid viscosity, flagellar geometric dimensions, and the bending stiffness of the flagellar hook. Here $C = \hat{\zeta}_{\perp} \cdot (L/3 + l/2)$. Based on the flagellar radius $a$ of \textit{E. coli}, the effective length of the flagellar bundle $L$, and the cell body half-length $l$, we have $\hat{\zeta}_{\perp} \approx 1.7$, yielding $C \sim 6.9 \, \mu\text{m}$.

Based on the experimental data in Fig. (\ref{fig:bacterial_wobbling_in_flow})a, the slope of the off-axis angle relative to the flow rate is $\approx 3.93 \times 10^9 \, \text{rad}/(\text{m}^3/\text{s})$. The relationship between the shear rate at $z = 1 \mu$m in the channel and the flow rate is $\dot{\gamma}(Q) = 6Q/(Wh^2)$. Therefore, the slope of the experimental curve is $k_{exp} = \mathcal{K} \cdot 6 \sin(\theta_0)/(Wh^2)$.
Given the channel parameters ($W = 600 \mu\text{m}, H = 100 \mu\text{m}$) and $\theta_0 = 10^\circ$, we obtain $\mathcal{K} \approx 0.023 \, \text{s}$. Given the fluid viscosity $\eta \approx 10^{-3} \, \text{Pa}\cdot\text{s}$ and flagellar length $L \approx 10 \mu\text{m}$, we can back-calculate the bending stiffness of the hook:
$K_h \approx 4.3 \times 10^{-17} \, \text{N}\cdot\text{m}/\text{rad}$.
This magnitude is comparable to the motor hook stiffness reported in the literature. Converting this using the relationship $K_h = EI / L_{hook}$ (where $L_{hook} \approx 55 \text{ nm}$) yields results consistent with the literature. Sen et al. obtained a bending stiffness of $\sim 10^{-29}$ to $\sim 1.8 \times 10^{-22}$ via in vitro electron microscopy, which was considered extremely soft \cite{sen2004elasticity}. Son et al. obtained $K_h \sim 4.0 \times 10^{-18}$ for \textit{Vibrio} \cite{kato2019structure}. Nord et al. obtained $K_h \sim 9 \times 10^{-19}$ to $5 \times 10^{-17}$ in single-hook torsion experiments, which is surprisingly consistent with our results; furthermore, they found that this stiffness increases with increasing load \cite{bergthorson2005particle}. Zhang et al. obtained $K_h \sim 10^{-18}$ from flow field measurements, a value slightly lower than our fitted experimental data \cite{zhang2023differential}.

The core conclusion of the study by Zhang et al. is that the bending stiffness of the hook is not constant; it increases with the load (flow field, torque). In the run state of \textit{E. coli}, where multiple flagella rotate in a bundle, the cooperative action of multiple hooks naturally exhibits a higher "equivalent bending stiffness" than a single hook. As stated by Mathijssen and Leishangthem \cite{mathijssen2019oscillatory, leishangthem2024thermodynamic}, due to the tilt, the shear gradient experienced by the flagellum is very large. In such a high-load environment, the hook is in a high-stress state, corresponding to the high-stiffness range measured by Nord et al. \cite{bergthorson2005particle}. As $\alpha$ increases, the wobble becomes more violent, which in turn generates a large periodic fluid resistance torque.
The effect of the near-wall flow field on bacteria is not just fluid drag, but rather the reshaping of their movement patterns by changing their geometric configuration (hook deformation). The research papers by Mathijssen et al. and Leishangthem et al. provide a solid foundation for "why bacteria tilt" \cite{mathijssen2019oscillatory, leishangthem2024thermodynamic}, while our experiment suggests "how the flow field regulates wobble via $\alpha$ under this tilt." As the flow rate $Q$ increases, $\alpha$ increases because the nose-down swimming posture puts the flagellum in a higher flow velocity layer, and the shear flow "pushes" the flagellum from behind toward the front. In the case of small flow rate $Q$, most bacteria are swimming upstream; this thrust enhances the original deviation tendency, thereby widening the off-axis angle, increasing $\theta_w$, and causing the bacterial swimming speed $V$ to decrease.

Once the off-axis angle $\alpha$ increases with the flow field, the rotation dynamics of the bacterium will undergo profound changes, thereby causing changes in the migration velocity. In the previous theoretical analysis, we considered swimming bacteria in a preset configuration, i.e., with a fixed off-axis angle $\alpha$ and constant flagellar motor output, and obtained the overall angular velocity $\bm{\Omega}$ and translational velocity $\mathbf{U}$. The overall total angular velocity $\bm{\Omega}$ is determined by the instantaneous torque balance. When $\alpha$ increases, the propulsion torque generated by the flagellum deviates further from the major axis, forcing the total rotation axis $\bm{\Omega}$ away from the cell body's major axis $\mathbf{e}_b$. According to the definition of the wobble angle $\theta_w$, the angle between the major axis $\mathbf{e}_b$ and the precession axis (the total rotation axis $\bm{\Omega}$) is approximately $\theta_w \approx \arccos(\mathbf{e}_b \cdot \bm{\Omega} / |\bm{\Omega}|)$. An increase in $\alpha$ directly drives an increase in $\theta_w$. As the flow rate $Q$ increases, the shear torque $T_{shear}$ increases, causing further bending deformation of the flagellar hook, i.e., increasing the off-axis angle $\alpha$. The increase in $\alpha$ forces the rotation axis to deviate from the major axis, making the wobble more intense, i.e., increasing the wobble angle $\theta_w$, which leads to a decrease in the projected component $V$, as observed in the speed drop in the experiment.

\subsubsection{Coupling with rheotaxis due to chiral flagellum}
However, the experimentally observed speed drop is much larger than the calculation based on RFT model with additional shear-induced torque bacteria. The chiral structure of the flagellar bundle, however, generates a transverse force in a shear flow field that is perpendicular to both the flow direction and the shear gradient. Bacteria near the wall exhibit a transverse drift across streamlines, a "side-swimming" phenomenon\cite{marcos2012bacterial, mathijssen2019oscillatory,jing2020chirality,dang2026}. This transverse swimming originates from the transverse force induced by flagellar chirality, which generates an additional chiral torque. When a left-handed helical flagellar bundle rotates in a flow field with a velocity gradient, the fluid resistance is unbalanced because different parts of the flagellar filament are in different flow velocity layers, thus producing a resultant force $\mathbf{F}_y$ in the $y$-direction perpendicular to the shear plane ($xz$-plane). This force acts on the flagellar bundle while the cell body's center of mass is at another position, thereby producing a deflection torque around the $z$-axis at hook position $\mathbf{r}_{h}$, i.e., $\mathbf{T}_{chiral} = \mathbf{r}_{h} \times \mathbf{F}_c$. This torque can drive the transverse rheotaxis of bacteria as well. As shown in Fig. \ref{fig:Alpha-Theta-flow}(\textbf{c}), bacterial wobbling changes due to hook deformation by torque balance between restoring torque and shear torque, chirality-induced lift torque. Simplified upstream bacteria has nose-down orientation $\theta_0$ with flagellar bundle initially off-axis of $\alpha_0$ tilted upward. Top view shows the chirality-induced lift force on the helical flagellum rotate the flagellum with the anchoring of the hook towards to the opposite of the flow vorticity.

In the previous discussion, we assumed that bacteria perfectly swim upstream against the flow on the wall, in which case the shear force acting on the bacterium and cell body is close to the axial direction of the body and flagellum. However, once drift swimming occurs, the orientation of the cell body and flagellum is close to being perpendicular to the flow field along the $y$-direction. In this case, the total shear force and shear torque are larger than those during perfectly upstream swimming. As shown in Fig. \ref{fig:bacterial_wobbling_in_flow}\textbf{b} and \textbf{c}, when completely perpendicular to the flow field, the wobble angle is largest, even much larger than the one during completely upstream swimming. The originally considered shear torque $T_{flow}$ will be amplified, thus doubling the wobble angle, allowing for a good match with the experiment. Near the wall at $z = 1 \mu$m, the torque generated by the shear flow field strongly depends on the direction of the cell body's major axis $\mathbf{e}_b$. In the completely upstream state, the angle between the cell body's major axis $\mathbf{e}_b$ and the velocity gradient direction ($z$-axis) is very small. In this case, the effective cross-sectional area of the "side-wind force" generated by the flow field on the cell body and flagellum is minimized. When the bacterium starts to deflect toward the $y$-direction due to the chiral force, the major axis enters a state perpendicular to the flow direction ($x$-axis). When the flagellar bundle (length 7 $\mu$m) is spread out in the $y$-direction, the shear torque experienced by its different segments is no longer just a simple axial component but the full cross-sectional resistance. The vertical drag coefficient $\zeta_{\perp}$ experienced by the flagellar bundle is about twice the parallel drag coefficient $\zeta_{\parallel}$. When the bacterium is deflected laterally, the flow field acts more on the $\zeta_{\perp}$ direction.

The chiral force formula is $F_c \propto \dot{\gamma} \cdot \omega_m \cdot L^2 \cdot \sin \theta_0$. This force is perpendicular to the flow plane and pulls the bacterium directly away from the axis. At this point, the hook, which originally maintained the forward progress, must not only resist axial shear but also the huge lateral torque generated by lateral deflection. The increase in off-axis angle $\alpha$ is no longer linear but increases as the deflection angle increases. This explains why the wobble is greatest during lateral swimming in the experiment. Define an amplification function $M(\Psi) = 1 + \sin^2(\Psi)$ that varies with the deflection angle $\Psi$. When $\Psi = 0$ (upstream), $M = 1$. When $\Psi = 90^\circ$ (lateral), $M = 2$. The relationship between the off-axis angle and the flow field can be approximately written as $\alpha(Q, \phi) \sim \alpha_0 + T_{shear}(Q) \cdot M(\phi) / K_h$. Increasing $F_c$ produces a stronger bending torque $T_{shear}$ thus increasing the flagellar hook deformation $\alpha$ and causing an increase in wobble. As seen from experimental Fig. \ref{fig:bacterial_wobbling_in_flow}, under the action of the chiral force, the azimuth angle becomes closer and closer to 90 degrees with the increase of the flow field, leading to higher wobbling angle (Fig. \ref{fig:bacterial_wobbling_in_flow})\textbf{c}. This means the amplification factor $M(\Psi)$ is approaching 2, with qualitative consistence for the wobble angle with and without this chiral force correction (Fig. S3). Note that, we find that compared with the experimental value, the wobble angle obtained by theoretical calculation after correction is still small.

\subsubsection{Trapping swimming on the wall}

Experimental observations above clearly reveal that the bacterial cell body can undergo continuous rolling using the contact point between the body and the wall as a fulcrum, as illustrated in Figs. \ref{fig:WobblingWithRolling}(\textbf{b,c}). The emergence of this rolling motion fundamentally reshapes the entire landscape of bacterial wobbling. Originating from the vorticity of the shear flow, this rolling is intrinsically coupled with the flow intensity and remains highly sensitive to the bacterium's center-of-mass height and orientation, significantly increasing the complexity of interpreting the wobbling dynamics.Consequently, the proximity to a solid interface poses substantial challenges for interpreting experimental data through simplified bulk RFT models. 

Near the channel surface, the no-slip boundary condition of the shear flow not only modulates the off-axis angle but also deterministically induces the aforementioned rolling motion of the bacterial body. This surface-driven component superimposes onto the intrinsic precessional dynamics of the wobbling gait, fundamentally altering the observed frequency response. The flow-induced increase in the off-axis angle $\alpha$ would theoretically predict a decay in the intrinsic wobbling frequency due to heightened hydrodynamic resistance (Fig. S1 in \textit{supplementary material}). However, our experimental measurements reveal a monotonic rise in frequency with flow strength shown in Figs. \ref{fig:WobblingWithRolling}(\textbf{b}). This result indicates that the shear-induced rolling near the wall effectively compensates for the frequency decay associated with structural reconfiguration, leading to the observed linear increase in the total wobbling frequency. 

In our experiments, bacteria are often "locked" on the wall plane at a given observation height (e.g., $z = 1 \mu$m). For the wobble itself, even if the $z$-translation is restricted, the body will still rotate in a conical fashion around the thrust axis due to the asymmetry of $\alpha$. This may explain why the wall provides a supporting force (preventing sinking) but cannot prevent the oscillation (wobble) caused by the torque.
Intuitively, if a bacterium performs nose-down parallel translation along the wall while the obliquely upward-pointing flagellum rotates at high speed and revolves around the precession axis, the flagellum's position relative to the cell body is sometimes obliquely above the body and sometimes obliquely below. The velocity and stress differences caused by the height difference between the flagellum and the body are not constant, and the resulting additional torque does not always increase the off-axis angle $\alpha$ by deforming the hook. However, in low Reynolds number fluids, the flagellum's rotation frequency ($\sim 100$ Hz) is much higher than the body's wobble frequency ($\sim 10$ Hz). In calculating hook deformation, the RFT theory typically takes the time-averaged torque over one flagellar rotation period. Although the instantaneous torque is oscillating, the steady-state deformation $\alpha$ of the hook depends mainly on the average component due to the hook's damping and the high fluid viscosity.

When an increase in $Q$ leads to an increase in $\alpha$, $\theta_w$ may deviate from a simple linear relationship with $Q$. Near the wall, because motion in the $z$-direction is suppressed, the resulting rotational torque is forced to couple to the $XY$ plane, making $\theta_w$ exhibit a "saturation" or a specific slope change. The true nonlinearity comes from the $\sin \alpha$ and $\cos \alpha$ terms included in the $\mathbf{A}_{tot}$ matrix. When $\alpha$ is small, $\sin \alpha \approx \alpha$. If the flow rate $Q$ is extended to $100$ nL/s, $\theta_w$ exhibits nonlinearity.

In our previous analysis, the translational velocity $\mathbf{U}$ is the relative velocity of the bacterium with respect to the local fluid, and the translational speed $U$ undergoes an offset effect due to "upstream" swimming. At $z = 1 \mu$m, the background flow velocity $u_{bg}$ increases linearly in the opposite direction with $Q$. If $Q$ is large enough and $U$ crosses zero, it means the bacterium cannot maintain upstreaming and starts to be washed away by the flow field. Here, wall effects are introduced through a correction term (wall resistance factor coefficient). Large speed drops are observed in the experiment, while the translational speed drop calculated by the theoretical model is minor; let's discuss possible reasons. There is a nonlinear enhancement of wall resistance: when a bacterium is nose-down near the ground, the distance between the head and the wall may enter the nanometer range. At this point, lubrication theory predicts that the resistance will diverge as $1/d$ (the inverse of the distance), far exceeding a factor of 2. In addition, an increase in flow rate leads to an increase in $\alpha$, and the axial component of the thrust is lost in proportion to $\cos \alpha$. When a small initial angle $\alpha$ is set, the change in this off-axis angle also leads to a non-significant drop in speed.

\section{Conclusion}
In summary, we investigated the complex interplay between external shear flow and the intrinsic swimming gait of flagellated bacteria near surfaces. 
By high-resolution microscopy, we elucidate the hydrodynamics governing the bacterial wobbling in realistic fluid environments.
Shear flow enhances the bacterial wobbling amplitude, with the wobbling angle, which reaches a plateau  at flow moderate rates. This intensification is potentially linked to the deformation of the flexible flagellar hook.
The chirality of the flagellar bundle introduces a transverse force that drives cross-stream migration. This reorientation toward a lateral swimming posture amplifies the effective shear torque, increasing the load on the hook when the bacterium is perpendicular to the flow. 
This explains the maximal wobbling observed during transverse drift.
Bacterial geometry acts as a critical regulator; shorter cells exhibit more pronounced wobbling due to lower rotational drag. 
Critically, we demonstrate that enhanced wobbling suppresses motility, reducing the net migration velocity by up to $15\%$. This decay results from the loss of axial thrust as energy is redirected into lateral oscillations and precession. Our results indicate that bacterial wobbling in flow is not a mere passive response but a dynamic gait adjustment dictated by the coupling of flow, morphology, and hook elasticity. These findings help to understand how microorganisms navigate complex shear environments.

\section*{ MATERIALS AND METHODS}

\subsection*{\label{sec:MatMethd}Bacteria culture}

 Cultures of \textit{Escherichia coli} (strain AW405) were prepared following a standard protocol. Briefly, cells were retrieved from a frozen tube stored at $-20~^\circ\mathrm{C}$ and inoculated into  $10~\mathrm{mL}$ of LB medium. The LB medium was prepared from $1.0~\%~(\mathrm{w/v})$ tryptone, $0.5~\%~(\mathrm{w/v})$ yeast extract, and $1.0~\%~(\mathrm{w/v})$ NaCl. Cultures were incubated overnight at $30~^\circ\mathrm{C}$ with shaking at 200 rpm (about 12 h). Subsequently, $100\ \mu\text{L}$ of culture was inoculated into fresh $10\ \text{mL}$ LB medium. Growth was continued under the same conditions until the optical density at $600~\mathrm{nm}$ (OD$_{600}$) reached $\sim 0.7$ (about 6 h). Cells were then centrifuged and subjected to fluorescent staining.

\textit{E. coli} (strain AW405) was fluorescently labeled with Alexa Fluor 532 (red, AAT Bioquest). After 6 hours of growth, the culture was centrifuged at $300~\mathrm{rpm}$, and the pellet was resuspended in deionized water to a total volume of $10~\mathrm{mL}$. The washing procedure was repeated until the supernatant was fully removed. A $500~\mathrm{\mu L}$ aliquot of the washed suspension was mixed with $10~\mathrm{\mu L}$ of dye and $25\mu \text{L}$ of sodium bicarbonate. The mixture was incubated in the dark on a shaker at 200 rpm for 30 min. After staining for 30 min, the bacterial suspension was centrifuged at 300 rpm and washed with deionized water to $10~\mathrm{mL}$ per cycle. Then, the bacterial pellet was resuspended in motility buffer (MB) to achieve a final optical density of approximately $\text{OD} \approx 0.3$ under the optical wavelength of $600~\mathrm{nm}$. The MB consisted of $1~\mathrm{M}$ sodium lactate, $100~\mathrm{mM}$ EDTA, $1~\mathrm{mM}$ L-methionine, and $0.1~\mathrm{M}$ potassium phosphate buffer, adjusted to pH 7.0. The bacterial suspension was subsequently introduced into the observation microfluidic channel for analysis.

\subsection*{Visualization}
Both fluorescent and nonfluorescent $\textit{E.coli}$ (strain AW405) were introduced into a polydimethylsiloxane (PDMS) microchannel via a microfluidic pump. The microchannel had a height of H=$100~\mu$m, a width of W=$600~\mu$m, and a length of L=$20~$mm (\ref{fig:setup} (\textbf{a})). Fluorescent cells were used for measuring the off-axis angle, whereas nonfluorescent cells were used for tracking bacterial wobbling under bright-field imaging.

To ensure consistent and comparable measurements, both types of cells were observed $1~\mathrm{\mu m}$  above the bottom surface. To observe the bacteria bodies and flagella clearly when applying different flow rates, we focused the inverted $60\times$ (NA= 1.2) microscope (Nikon Ti2-E, Japan) and high-speed camera (Hamamatsu, ORCA-Flash4.0 V3), which covers a field of view \( 225 \times 225 \ \mu \text{m}^2 \). To ignore the influence of the Poiseuille flow generated by the side walls, the field of view was fixed at a distance of $200\mu \text{m}$ from the channel walls. Due to prolonged fluorescent exposure, which reduces the brightness of stained bacteria by approximately 30\%, a fresh batch of bacteria was supplied for observation after each video capture (approximately 10 s).

\subsection*{Tracking and analysis}
The instantaneous orientation and position of individual bacteria were recorded using a custom-written tracking algorithm to analyze their swimming trajectories (\ref{fig:setup} (\textbf{b})). Due to the misalignment between the bacterial body and the flagellar bundle, the body orientation exhibits periodic oscillations. The wobbling angle ($\theta_w$) was defined as half of the maximum oscillation amplitude, obtained by tracking the instantaneous orientation of the cell body over time (\ref{fig:setup} (\textbf{c})). To reduce the influence of noise, such as circular motion and other sources of fluctuations, the orientation time series $\Psi$ was processed using a discrete Fourier transform (DFT) to identify the main oscillation frequency, followed by filtering of unwanted frequency components and an inverse Fourier transform to accurately extract the oscillation amplitude and frequency.

In addition, to characterize the deviation of the cell body from the flagellar bundle, the off-axis angle $\alpha$ was directly measured as the angle between the bacterial body orientation and the opposite extension of the bundle axis.

\begin{acknowledgments}
The authors acknowledge the founding support from National Natural Science Foundation of China (No.12474197), the Shaanxi Academy of Fundamental Sciences (Mathematics, Physics No.23JSY024) and  Innovation Capability Support Program of Shaanxi(Program No.2025ZC-KJXX-51).
\end{acknowledgments}

\bibliography{Ref}
\end{document}